\documentclass[preprint,fleqn,usenatbib]{aastex63}
\usepackage{color}
\usepackage{lineno}
\usepackage[title]{appendix}

\begin{document}
 
\shorttitle{Period Changes of 14,127 CEBs in the Galactic Bulge}
\shortauthors{Hong et al.}

\title{Period Changes of 14,127 Contact Eclipsing Binaries in the Galactic Bulge}

\correspondingauthor{Kyeongsoo Hong}

%
\email{kyeongsoo76@gmail.com}

\author[0000-0002-8692-2588]{Kyeongsoo Hong}
\affiliation{Institute for Astrophysics, Chungbuk National University, Chungdae-ro 1, Seowon-Gu, Cheongju 28644, Republic of Korea}

\author[0000-0002-5739-9804]{Jae Woo Lee}
\affiliation{Korea Astronomy and Space Science Institute, Daejeon 34055, Republic of Korea}

\author[0000-0001-9339-4456]{Jang-Ho Park}
\affiliation{Korea Astronomy and Space Science Institute, Daejeon 34055, Republic of Korea}

\author[0000-0002-6687-6318]{Hye-Young Kim}
\affiliation{Department of Astronomy and Space Science, Chungbuk National University, Cheongju 28644, Republic of Korea}

\author[0000-0003-0043-3925]{Chung-Uk Lee}
\affiliation{Korea Astronomy and Space Science Institute, Daejeon 34055, Republic of Korea}

\author[0000-0001-8263-1006]{Hyoun-Woo Kim}
\affiliation{Department of Astronomy and Space Science, Chungbuk National University, Cheongju 28644, Republic of Korea}

\author[0000-0000-0000-0000]{Dong-Jin Kim}
\affiliation{Korea Astronomy and Space Science Institute, Optical Division, 776 Daedeok-daero, Yuseong-gu, Daejeon 34055, Republic of Korea}

\author[0000-0002-2641-9964]{Cheongho Han}
\affiliation{Department of Physics, Chungbuk National University, Cheongju 28644, Republic of Korea}

\begin{abstract}

We present the orbital period variations of 14,127 contact eclipsing binaries (CEBs) based on the OGLE-III\&IV observations in the Galactic bulge. New times of minimum lights for the CEBs were derived by binary modeling for the full seasonal light curves, which were made from survey observations at an interval of 1 year.
The orbital period changes of the systems were classified based on the statistical inference, multiple-hypothesis testing error measure, and visual inspection of the eclipse timing diagrams. As the results, we identified 13,716 CEBs with a parabola, 307 CEBs with a sinusoid, and 104 CEBs with the two variations. 
The period distributions of the inner close binaries and the outer companions were in the ranges of $0.235-0.990$ days and $5.0-14.0$ years, respectively. In our sample of 13,820 CEBs showing a parabolic variation, the highest decreasing and increasing period rates were determined to be $\dot P=-1.38\pm0.06\times10^{-5}$ day year$^{-1}$ for OGLE-BLG-ECL-169991 and $\dot P=+8.99\pm0.44\times10^{-6}$ day year$^{-1}$ for OGLE-BLG-ECL-189805, respectively. The secular period change rates were distributed almost symmetrically around zero, and most of them lie within $\dot P=\pm5.0\times10^{-6}$ day year$^{-1}$.

\end{abstract}


\section{Introduction}

Generally, contact eclipsing binaries (CEBs) consist of two dwarf stars surrounded by a common convective envelope. Most of these systems have spectral classes from F to K, and orbital periods between 0.2 and 1.0 days (Maceroni \& van’t Veer 1996; Paczy\'nski et al. 2006). The light curves of these binaries show similar eclipse depths because they have almost the same effective temperatures due to the mass and energy transfers between the components (Lucy 1968; Webbink 2003). A good review of the possible formation and evolution processes of close binary system have been provided by Webbink (1976), St\c epie\'n (2006), and Eggleton (2012).

The investigation of orbital period changes of the CEBs has contributed to the understanding of binary evolution.
In general, the eclipse timing variations are represented by one or a combination of parabolic or cyclical variations created by various effects, as follows:
(i) the light-travel-time effect (LTTE) due to additional companions (Irwin 1952, 1959) or period oscillations because of magnetic activity cycles (Applegate 1992; Lanza et al. 1998) show cyclical period changes,
(ii) the angular momentum loss (AML) due to a gravitational radiation or magnetic braking (Paczy\'nski 1967; van't Veer 1979; Mochnacki 1981; Rappaport et a. 1983; van’t Veer \& Maceroni 1989) indicate downward parabolic changes created by the period decrease,
(iii) mass ejection from the binary system (Prendergast \& Taam 1974) demonstrates a upward parabolic variation induced by the period increase; and
(iv) mass transfers between the binary components (Lucy 1976; Webbink 1976, 2003) can show upward or downward variations.

Since 1992, the Optical Gravitational Lensing Experiment (OGLE) project has carried out long-term wide-field survey observations, searching for dark and unseen matter using the microlensing phenomena (Udalski et al. 1992). Recently, Soszy\'nski et al. (2016) published a list of 450,598 eclipsing and ellipsoidal binary systems (86,560 CEBs, 338,633 semi-detached and detached EBs, and 25,405 non-eclipsing ellipsoidal variables) based on the OGLE-II, OGLE-III, and OGLE-IV observations in the Galactic Bulge. The long-term homogeneous data of eclipsing binaries allows us to study variations in their orbital period.
Kubiak et al. (2006) reported secular period variations for 569 CEBs with a short period ($P \leq 1$ day) from the OGLE data between 1992 and 2005. Pietrukowicz et al. (2017) identified 56, 52, and 35 systems with increasing, decreasing, and cyclic period variations, respectively, by searching for potential stellar mergers in a sample of 22,462 short period EBs ($P \leq 4$ day) in the Galactic bulge using OGLE observations from 1992-2015. Hajdu et al. (2019) reported LTTE solutions for 992 hierarchical multiple candidates, which were determined by analyzing the eclipse timing diagrams of about 80,000 EBs with a period shorter than 6 days.

Luminous Red novae (LRNe) are very important targets when investigating the evolution of binary systems, and have been related to common-envelope evolution (Soker \& Tylenda 2003; Kulkarni et al. 2007; Tylenda et al. 2011; Ivanova et al. 2013; Nandez et al. 2014; Pastorello et al. 2019; Howitt et al. 2020). Only four of these LRNe have been noted in our Galaxy: V4332 Sgr (Hayashi et al. 1994), V838 Mon (Brown et al. 2002), V1309 Sco (Nakano et al. 2008), and OGLE-2002-BLG-360 (Tylenda et al. 2013). The LRNe are known to be the result of the merger of the cores of contact binaries. They have a total energy in the range of $10^{45}-10^{48}$ ergs (Kashi 2018), which have intermediate luminosities between classical novae and super-novae, and relatively long outbursts. Kochanek et al. (2014) estimated that these transients are produced about once per decade. According to the suggestion by Ferreira et al. (2019), the magnitude of LRNe V1309 Sco has nearly stabilized during about 9 years after the outburst, and the system is likely to be in a blue straggler state. Recently, the Contact Binaries Towards Merging (CoBiToM) Project was performed to study stellar evolution before the merging processes by Gazeas et al. (2021).

The aim of this study was to search for potential stellar mergers through the eclipse timing analysis of a total of 86,560 CEBs classified by Soszy\'nski et al. (2016). The homogeneous data obtained from the OGLE-III\&IV surveys provide long-term stable and uniform photometry for the objects, over a long period of 15 years. Section 2 provides the information about the OGLE data used in this study. We briefly describe the determination of the times of minimum lights for the eclipse timing diagrams in Section 3. Section 4 presents the study of the orbital period variations of the selected CEBs. Finally, we present the summary and discussions of our results in Section 5.

\section{Data collections}

For this study, we collected photometric data for 86,560 CEBs from the OGLE-III\&IV archive data from 2001 to 2015 for the Galactic bulge fields. The OGLE observations were carried out using the 1.3-m Warsaw telescope at the Las Campanas Observatory in Chile. The telescope is equipped with an eight-CCD mosaic camera covering a 0.35 deg$^{2}$ field of view during the OGLE-III and with a 32 chip mosaic camera which covers a 1.4 deg$^{2}$ field of view for the OGLE-IV. The sky coverages were nearly 92 deg$^{2}$ for the OGLE-III in 2001$-$2009 and 182 deg$^{2}$ for the OGLE-IV in 2010$-$2015. In this study, only light curves in $I$ band were used because the surveys were obtained mostly in $I$ band, occasionally in $V$ band. The overall specification and reductions of the OGLE observations were provided by Udalski et al. (2008 for OGLE-III; 2015 for OGLE-IV).

\section{Times of minimum light}

For the orbital period studies of the CEBs, we used the observed minus computed ($O-C$) diagram, which is a powerful diagnostic tool that can measure very subtle changes in the orbital period. The eclipse timings for the $O-C$ diagram are usually determined using consecutive observations during primary or secondary eclipses. However the CEBs were observed only one to about thirty points per night for lowest and highest cadence fields by the OGLE-III\&IV surveys in the Galactic bulge. The number of data points in the $I$ band during the OGLE-III observation periods was collected from about 35 to 2,500 points per field, while the OGLE-IV observations were obtained about 150 to 17,000 measurements per field for their observing periods. More detailed information of the observation cadence for the OGLE surveys was presented in the papers of Udalski et al. (2003 for OGLE-III; 2015 for OGLE-IV).

In order to determine the times of minimum lights from the OGLE data, we adopted the method of determining the primary eclipse timing from the synthetic light curve, as applied in the papers of Hong et al. (2015, 2016, and 2019). First, full light curves with an interval of 1 year were constructed from the OGLE observations. Then, we analyzed the light curves using the binary modeling and obtained the primary eclipse timings from the binary solutions. For the binary modeling, we used the Wilson-Devinney program (Wilson \& Devinney 1971; Van Hamme \& Wilson 2007, hereafter W-D), which is an excellent tool for solving a binary star light curve. The light curve solutions by the W-D program can provide times of minimum light without the impact of spot activity (Maceroni \& van't Veer 1994; Lee et al. 2014).

\section{Eclipse timing variations}

In the eclipse timing analysis, times of minima with a significantly large error can increase the uncertainty of period changes.
Therefore we excluded the minimum times that had an uncertainty of more than three times the mean error (0.0025 days), which was calculated using the errors of all minimum times obtained from the individual light curves of 86,560 CEBs. Then the CEBs with eclipse timings with more than six points were selected by considering a variable number of the following equations in this section, to prevent incorrect results due to insufficient data. Consequently, we selected 50,824 of 86,560 CEBs with an orbital period range of $0.166-2.07$ days.

First of all, the mean light ephemeris for all selected systems was calculated introducing all minimum times into the following linear ephemeris: 
\begin{equation}
 C_1 = T_0 + PE. 
\end{equation}

Then, the selected systems were classified into three types (parabolic, sinusoidal, and parabolic $plus$ sinusoidal variations) based on the approach as presented below.
\begin{verse}
1. In order to classify the parabolic variations in the $O-C$ diagrams, we adopted the Bayes factor for a statistical inference (Kass \& Raftery 1995; Wagenmakers 2007) and the controlling false discovery rate (FDR) for a multiple-hypothesis testing error measure (Benjamini \& Hochberg 1995) based on the R software environment\footnote{\url{http://www.rproject.org}}, which is an open-source free statistical environment developed under the GNU GPL (Ihaka \& Gentleman 1996). 
The Bayesian information criterion (BIC) is given by 
\begin{displaymath}
BIC(H_i)=-2{\rm log}L_i+k_i{\rm log}~n,
\end{displaymath}
where $L_i$ and $k_i$ are the maximum likelihood and the number of free parameters for model $H_i$, respectively, and $n$ is the number of minimum times. The Bayes factor is given by 
\begin{displaymath}
BF_{10} = \exp \left( \frac{BIC(H_0)-BIC(H_1)}{2} \right).
\end{displaymath}
In this case, the hypothesis $H_0$ and $H_1$ represent the linear and parabolic regressions for individual CEBs, respectively.
We selected the models with the Bayes factor of $BF_{10} >10$ (cf. the scales of evidence for model selection are proposed in the paper of Kass \& Raftery 1995).
The controlling FDR is given by  
\begin{displaymath}
P_i \leq \frac{i}{m}\alpha,
\end{displaymath}
where $P_i$ is the ordered p-values and $m$ is the total number of hypotheses. We strict set the $\alpha$ to 0.01 for a false positive to be less than 1\%. 
In the case of the p-value of $H_0$ less than $H_1$, the hypothesis is rejected.
Applying the criterion, we identified 13,716 CEBs exhibiting parabolic variations.
\end{verse}
\begin{verse}
2. The sinusoidal and parabolic $plus$ sinusoidal variations in the $O-C$ diagrams of 50,824 CEBs were selected by visual inspection because it is difficult to solve the nonlinear least-squares fitting for these two models. Their fitting process requires good starting values of amplitude, angular frequency, and phase.
In the procedure, we excluded the systems with the period modulation shorter than 5 years by considering the one-year interval of eclipse timing data.
Then the $O-C$ diagrams of the selected systems were fitted according to the procedures described in Section 4.2 and 4.3, and verified using the Bayes factor and FDR.
As a result, we found the 307 CEBs with a sinusoid and 104 CEBs with two variations.
Note, all of the individual eclipse timings for 14,127 CEBs are listed in the machine-readable data table.
\end{verse}


\subsection{Parabolic variations}

The parabolic variations of CEBs can be explained by angular momentum loss due to a magnetic stellar wind, a mass transfer from the more massive to the less one, or a part of the periodic variation by the Applegate (1992) mechanism or LTTE. All of the eclipse timings of the 13,716 CEBs showing parabolic variations and the 104 CEBs with parabolic $plus$ sinusoidal variations were applied by using a least-squares parabolic fitting as follows,
\begin{equation}
 C_2 = T_0 + PE + AE^2.
\end{equation}
In order to check whether our processes were appropriate, the period change rates were compared with the results of 45 CEBs overlapped with ours in the paper by Pietrukowicz et al. (2017), where a total of 23 and 22 CEBs had decreased and increased period changes, respectively. 
The resulting period change rates ($\dot P$) for the overlapping CEBs are listed in Table 1, where the value of $\dot P$ is calculated from the relation $\dot P=2A/P$.
As one can see in Figure 1, the results are in satisfactory agreement with each other within their uncertainties.

Among the 13,716 CEBs, 6,263 and 7,453 systems were classified with decreasing and increasing period variations, respectively. 
Of the 104 CEBs with two variations, 69 and 35 systems were identified with decreasing and increasing period variations, respectively, which will be discussed in the following Subsection 4.3.
The resulting parabolic ephemeris for the 13,716 CEBs are provided in Table 2, which are sorted from the highest decreasing rate.

In our samples, the systems OGLE-BLG-ECL-169991 and OGLE-BLG-ECL-052921 had the highest period decreasing rates of $\dot P=-1.38\pm0.06\times10^{-5}$ day year$^{-1}$ and $\dot P=-9.74\pm0.54\times10^{-6}$ day year$^{-1}$, respectively, and the systems OGLE-BLG-ECL-189805 and OGLE-BLG-ECL-288708 had the highest period increasing rates of $\dot P=+8.99\pm0.44\times10^{-6}$ day year$^{-1}$ and $\dot P=+8.29\pm0.66\times10^{-6}$ day year$^{-1}$, respectively. The eclipse timing diagrams for the two CEBs with the highest period decreasing rate and for two CEBs with the highest period increasing rate are displayed in the upper and lower panels of Figure 2, respectively. The $I$ light curves phased with the linear and quadratic ephemeris for the four CEBs are plotted based on the OGLE-III and OGLE-IV observations in the left and right panels of Figure 3, respectively. In the left panels, the linear ephemeris taken from the OGLE EBs catalog presented by Soszy\'nski et al. (2016) were used, while the parabolic ephemeris derived from Equation (2) were applied in the right panels. As one can see in each panel of this figure, the scatters around the light curves were remarkably diminished after applying the parabolic ephemeris.

\subsection{Sinusoidal variations}

The sinusoidal variations in the $O-C$ diagram can be caused by LTTE due to an additional companion in a binary system (Irwin 1952, 1959) or period oscillations because of magnetic activity cycles (Applegate 1992; Lanza et al. 1998). To analyze the $O-C$ diagrams of 307 CEBs with a sinusoidal period variation, all times of minimum of each EB were fitted using a sine curve, as follows:
\begin{equation}
 C_3 = T_0 + PE + K \sin(\omega E + \omega_0 ). 
\end{equation}
The sinusoidal ephemeris gave an acceptable fit to the eclipse timings for 307 CEBs. In all calculations, the parameters in Equation (3) were solved based on the Levenberg-Marquart (hereafter LM; Press et al. 1992) technique. The basic parameters with fitting results for 307 CEBs are listed in Table 3, where the outer period ($P_2$) of the third companion is derived from the relation $P_2=2\pi P / \omega$.
The $O-C$ diagrams for 307 CEBs with the sinusoidal ephemeris are displayed in Figure 4. For comparison, the periods of the triple system candidates given in the papers of Pietrukowicz et al. (2017) and Hajdu et al. (2019) are presented in the last column of Table 3. A total of 22 of our and those systems were overlapped.

\subsection{Parabolic $plus$ sinusoidal variations}

As mentioned in Section 4.1, the eclipse timing diagrams of 104 CEBs show the parabola $plus$ sinusoidal variations. Therefore the $O-C$ diagrams were fitted using the quadratic ephemeris together with a sine curve, as follows:
\begin{equation}
 C_4 = T_0 + PE + AE^2 + K \sin(\omega E + \omega_0 ). 
\end{equation}
The basic parameters of 104 systems with parabolic and sinusoidal fitting results are listed in Table 4. In the panels of Figure 5, the blue dashed and red solid lines represent the parabolic term and the full ephemeris, respectively.

\subsection{Tertiary periods of triple system candidates}

Although the period changes of the CEBs with a sinusoidal variation can be explained by both magnetic activity cycles and LTTE, we only assumed the period variation of the systems by LTTE in this study. According to our results, the distributions of the inner ($P_1$; the period of close binary) and outer periods ($P_2$; the period of the third companion) for 411 triple system candidates were in the ranges of $0.235-0.990$ days and $5.0-14.0$ years, respectively. In this study, we first found 379 triple candidates, excepting the 32 EBs investigated previously in the papers of Pietrukowicz et al. (2017) and Hajdu et al. (2019). A histogram of their outer periods ($P_2$) is presented in Figure 6, where most values are located in the range of $3,400-3,800$ days. The mean of the outer periods in our results was determined to be $P_2\sim3200$ days. 
This is longer than the values of $P_2\sim2000$ days reported in the paper by Hajdu et al. (2019) because 
the observation period used in the study is longer than theirs. In the case of the real outer period that is significantly longer than the observation length, the calculated value would be shorter or closer period than the duration of the observation time span (cf. Hajdu et al. 2019).

Figure 7 shows the locations of the triple candidates (open circles) on the $P_1$ versus $P_2$ plane, where the grey circles represent the triple (or multiple) system candidates identified by Hajdu et al. (2019). In the figure, the horizontal and sloped blue lines represent the limits of the amplitudes with 50 seconds of the LTTE and dynamical effects (cf. Borkovits et al. 2016), respectively. As one can see in the figure, almost all of the triple candidates in this study are located in the region of higher LTTE contributions ($A_{\rm LTTE}$) than the dynamical effect ($A_{\rm dyn}$).

\section{Summary and Discussion}

In this study, we present the results of an eclipse timing analysis of 14,127 CEBs based on OGLE-III\&IV observations between 2001 and 2015 in the Galactic  bulge.
Among the 14,127 CEBs, 13,716 CEBs with parabolic variations were classified by the Bayes factor for a statistical inference (Kass \& Raftery 1995; Wagenmakers 2007) and the controlling false discovery rate (FDR) for a multiple-hypothesis testing error measure (Benjamini \& Hochberg 1995).
The 307 CEBs with a sinusoid, and 104 CEBs with the two variations were also identified by visual inspection and verified based on the Bayes factor and FDR.
Of the CEBs with a parabolic variation, 6,332 and 7,488 systems were classified as having decreasing and increasing period variations, respectively.
Of a total of 411 CEBs showing periodic variations, their inner and outer orbital periods are in the ranges of $0.235-0.990$ days and $5.0-14.0$ years, respectively. These were mainly dominated by the LTTE contribution rather than by the dynamical effect (cf. Figure 7).

Kubiak et al. (2006) suggested that the period change rates of the CEBs show around zero and up to $\pm2.3\times10^{-7}$ day year$^{-1}$ based on the distribution of the period variations for 134 CEBs in the Galactic bulge.
They did not find CEBs with a period change rate greater than $\pm5\times10^{-6}$ day year$^{-1}$. In this study, the number of CEBs with decreasing and increasing periods were 6,332 and 7,488 objects, respectively. A histogram of the orbital periods and the diagram of the period change rate versus the orbital periods are presented in the upper and lower panels of Figure 8, respectively. As shown in the upper panel of the figure, most CEBs have orbital periods in the range of $0.2 \leq P \leq$ 0.6 days, and the peak of the decreasing period shows slightly shorter orbital periods than the peak of the increasing period. In the lower panel of the figure, their period change rates are distributed almost symmetrically around zero, and they were mostly located within the range of $\dot P=\pm5.0\times10^{-6}$ day year$^{-1}$. In our sample, the highest decreasing and increasing period rates were determined to be $\dot P=-1.38\pm0.06\times10^{-5}$ day year$^{-1}$ for OGLE-BLG-ECL-169991 and $\dot P=+8.99\pm0.44\times10^{-6}$ day year$^{-1}$ for OGLE-BLG-ECL-189805, respectively. The period change rates in particular seem to be decreased on both sides of the peak around the orbital period of 0.4 days.

The variation in the period of the precursor to the LRNe V1309 Sco, with an orbital period of 1.4 days, decreased exponentially, from $\dot{P}=-9.64\times10^{-4}$ day year$^{-1}$ in 2002 to $\dot{P}=-3.93\times10^{-3}$ day year$^{-1}$ in 2007 before the eruption, and the shape of their light curves was transformed from a double to single wave (Tylenda et al. 2011). Molnar et al. (2017) suggested that the contact binary KIC 9832227 could be considered as a merger candidate based on a tentative exponential period decrease. However the system may not be a merger candidate according in a recent study (Socia et al. 2018; Kovacs et al. 2019).
Pietrukowicz et al. (2017) presented three potential merger candidates, EBs with relatively long orbital periods of more than $P>1.0$ days and a relatively high decreasing period rate of more than $\dot{P} >-1.0\times10^{-5}$ day year$^{-1}$. According to our results (cf. Table 2), a total of 70 systems have a period decreasing rate higher than $\dot{P}=-3.65\times10^{-6}$ day year$^{-1}$, which is a necessary criterion to be a serious merger candidate as presented by Molnar et al. (2017). Although the decreasing period of the systems can be interpreted as a part of the periodic variation due to the Applegate (1992) mechanism or LTTE, we carefully propose that the CEBs may be potential merger candidates.

\acknowledgments{Acknowledgements}
We are grateful to the anonymous statistics editor, referee, and Dr. Seung-Lee Kim for very helpful comments and suggestions that have helped us to improve the quality of the manuscript.
We thank the OGLE teams for all of the observations.
This research was supported by Basic Science Research Program through the NRF funded by the Ministry of Education (2019R1I1A1A01056776), and by grants of the National Research Foundation of Korea (2019R1A2C2085965 and 2020R1A4A2002885).


\clearpage
\begin{figure*}[!ht]
\vspace*{0pt}
\begin{center}
\begin{tabular}{c}
\hspace*{0pt}\includegraphics[width=0.45\columnwidth]{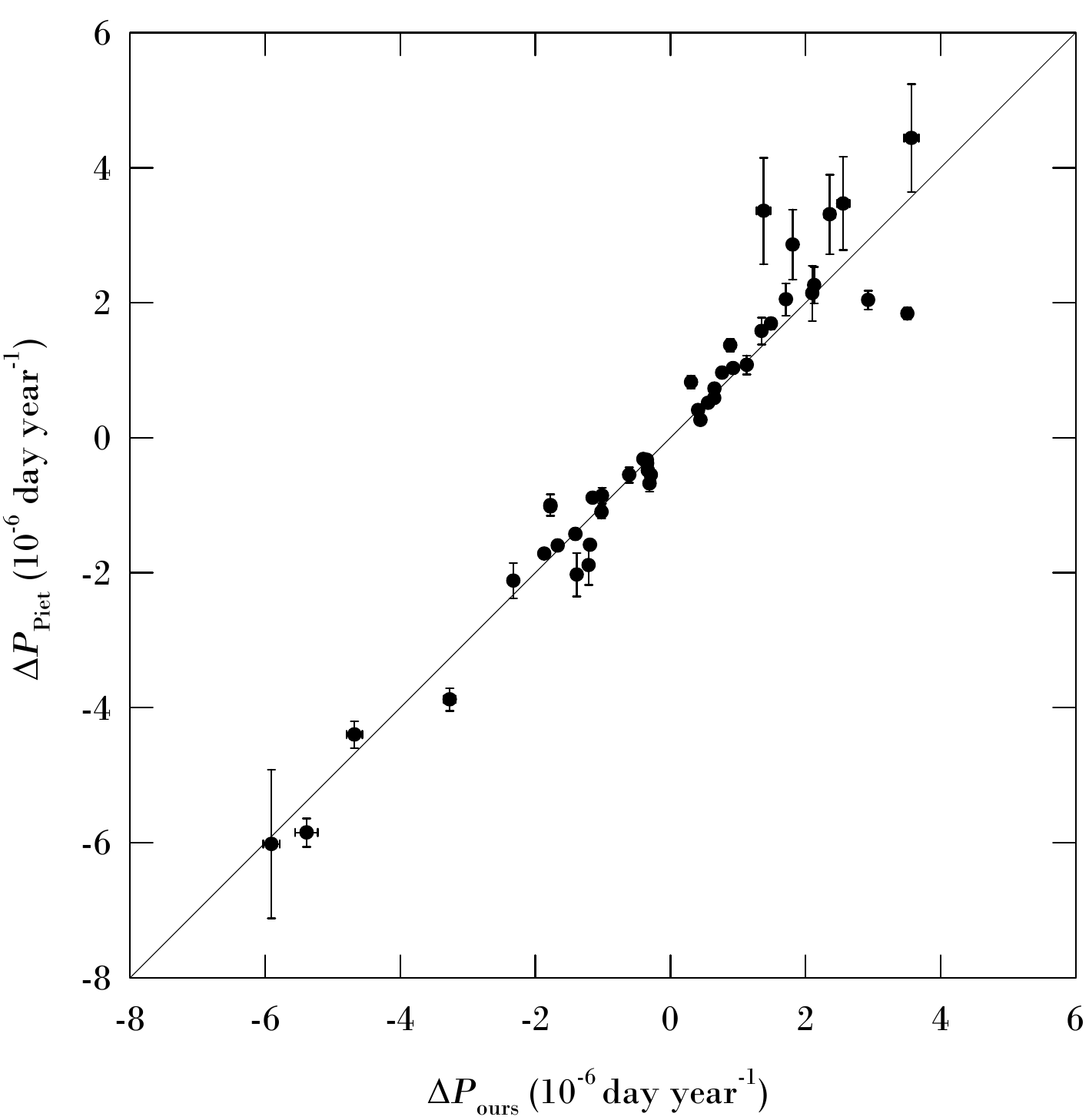}
\end{tabular}
\caption{Comparison of the period change rates between ours and those of Pietrukowicz et al. (2017).}
\end{center}
\vspace*{0pt}
\end{figure*}


\begin{figure*}[!ht]
\vspace*{0pt}
\begin{center}
\begin{tabular}{cc}
\hspace*{0pt}\includegraphics[width=0.65\columnwidth]{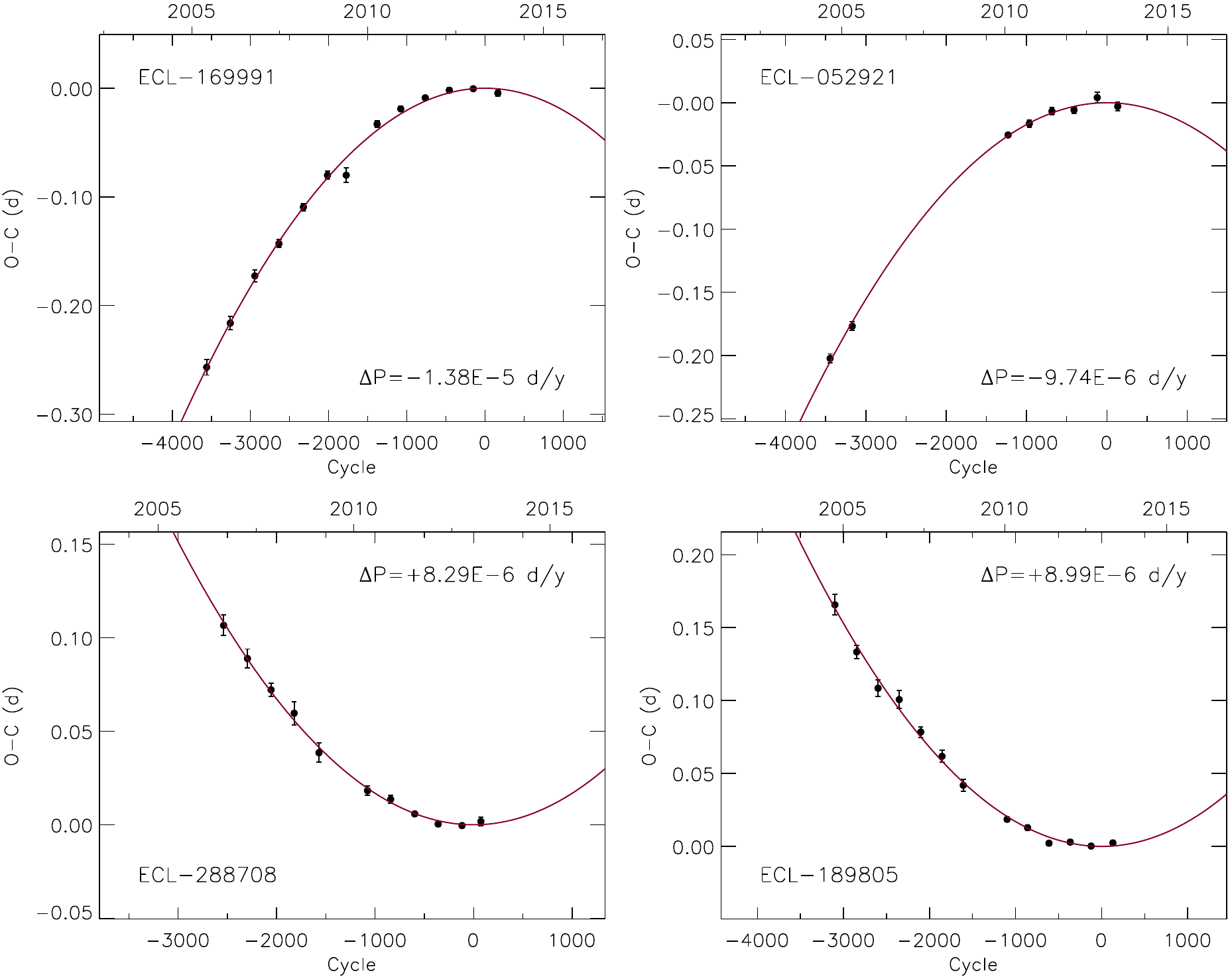} 
\end{tabular}
\caption{$O-C$ diagrams of four CEBs, which have the highest decreasing (upper panels) and increasing (lower panels) period change rates in our samples, respectively.}
\end{center}
\vspace*{0pt}
\end{figure*}

\clearpage
\begin{figure*}[!ht]
\vspace*{0pt}
\begin{center}
\begin{tabular}{cc}
\hspace*{0pt}\includegraphics[width=0.8\columnwidth]{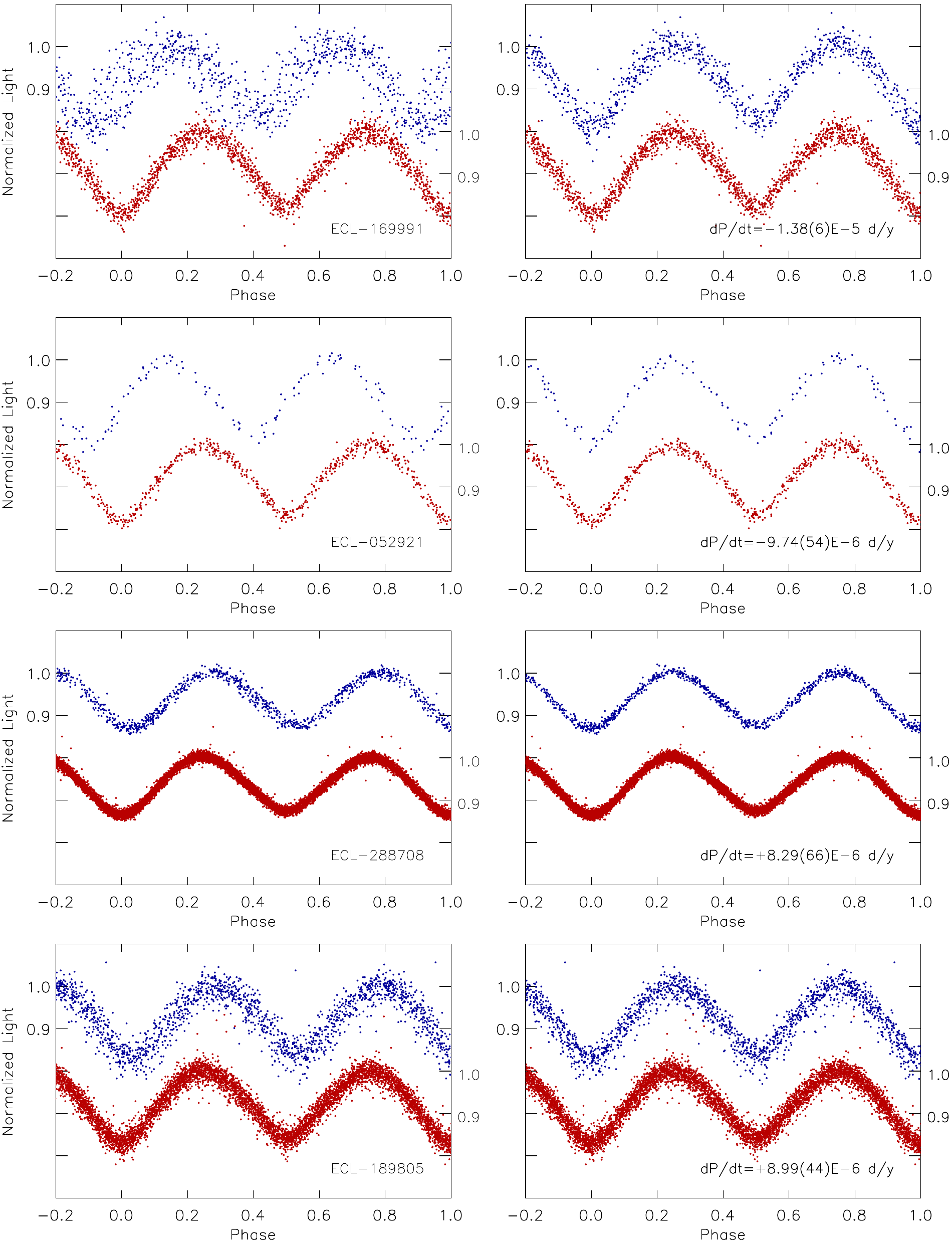}  
\end{tabular}
\caption{$I$ light curves of four CEBs. The blue and red circles represent the OGLE-III and OGLE-IV observations, respectively. In the left panels, the light curves were plotted using the linear ephemeris taken from the OGLE EBs catalog of Soszy\'nski et al. (2016).
In the right panels, the light curves were phased with the quadratic ephemeris determined in this study.}
\end{center}
\vspace*{0pt}
\end{figure*}

\clearpage
\begin{figure*}[!ht]
\vspace*{0pt}
\begin{center}
\begin{tabular}{c}
\hspace*{0pt}\includegraphics[width=0.9\columnwidth]{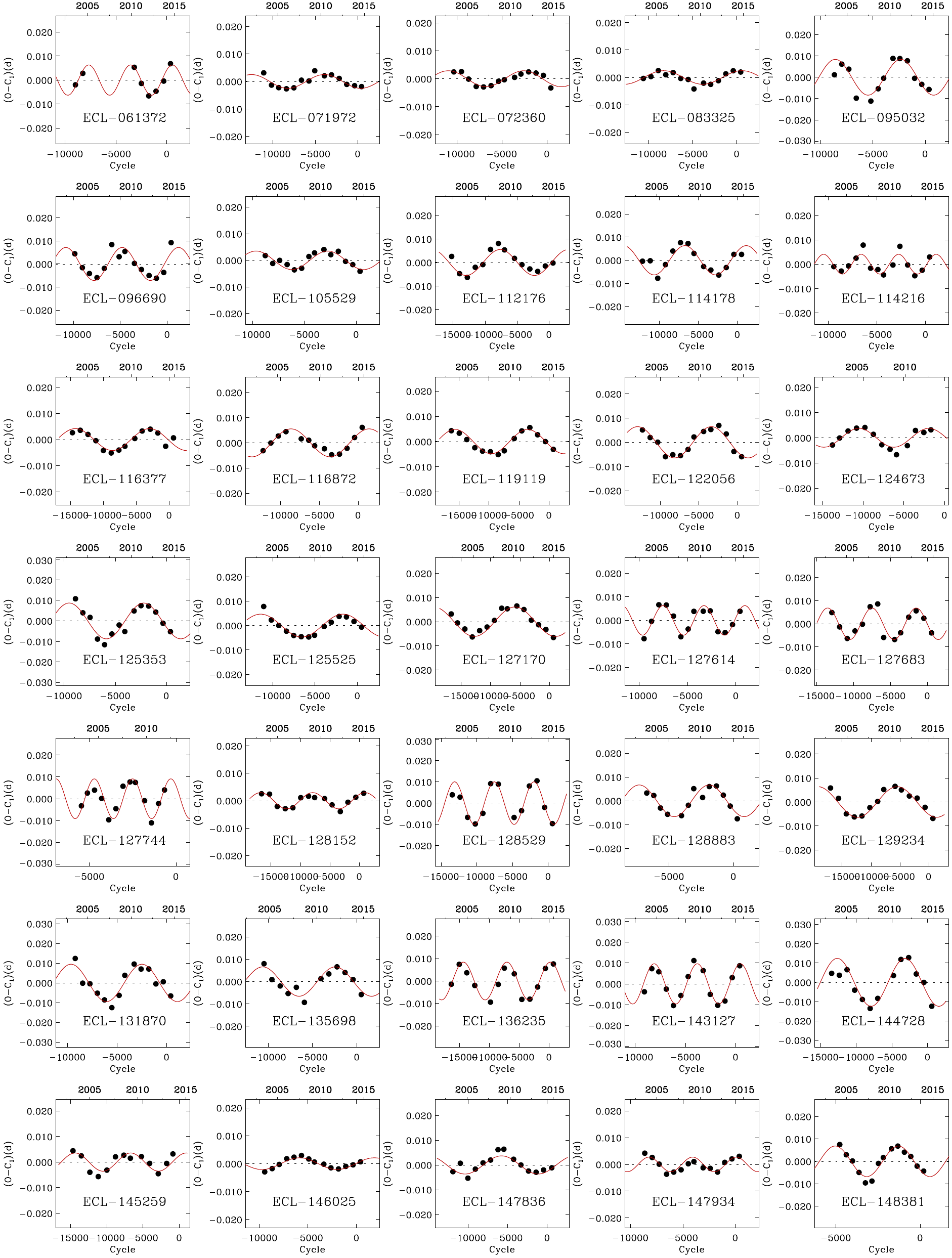} 
\end{tabular}
\caption{Samples for the eclipse timing diagrams of the 307 CEBs with a sinusoidal variation. In the panels, the red curves represent the sinusoidal term of the equation (3).
 (The full figures for the 307 CEBs with a sinusoidal variation can be obtained in the electronic edition of the paper.)
 }
\end{center}
\vspace*{0pt}
\end{figure*}

\begin{figure*}[!ht]
\vspace*{0pt}
\begin{center}
\begin{tabular}{c}
\hspace*{0pt}\includegraphics[width=0.9\columnwidth]{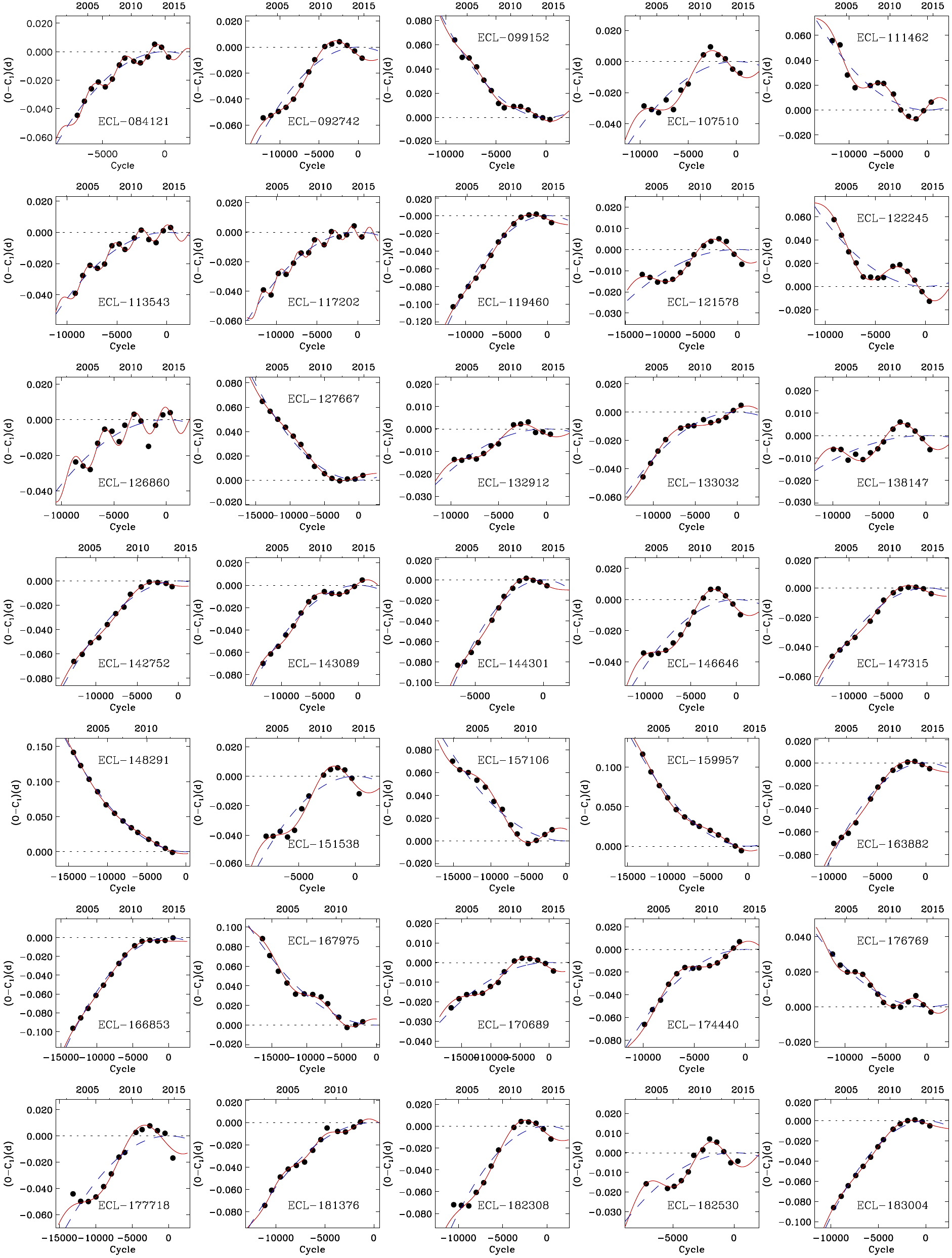} 
\end{tabular}
\caption{Samples for the eclipse timing diagrams of the 104 CEBs with parabolic $plus$ sinusoidal variations. 
In the panels, the blue dashed and red solid lines represent the parabolic terms and the full contributions, respectively.
 (The full figures for the 104 CEBs with parabolic $plus$ sinusoidal variations can be obtained in the electronic edition of the paper.)
}
\end{center}
\vspace*{0pt}
\end{figure*}


\begin{figure*}[!ht]
\vspace*{0pt}
\begin{center}
\begin{tabular}{c}
\hspace*{0pt}\includegraphics[width=0.58\columnwidth]{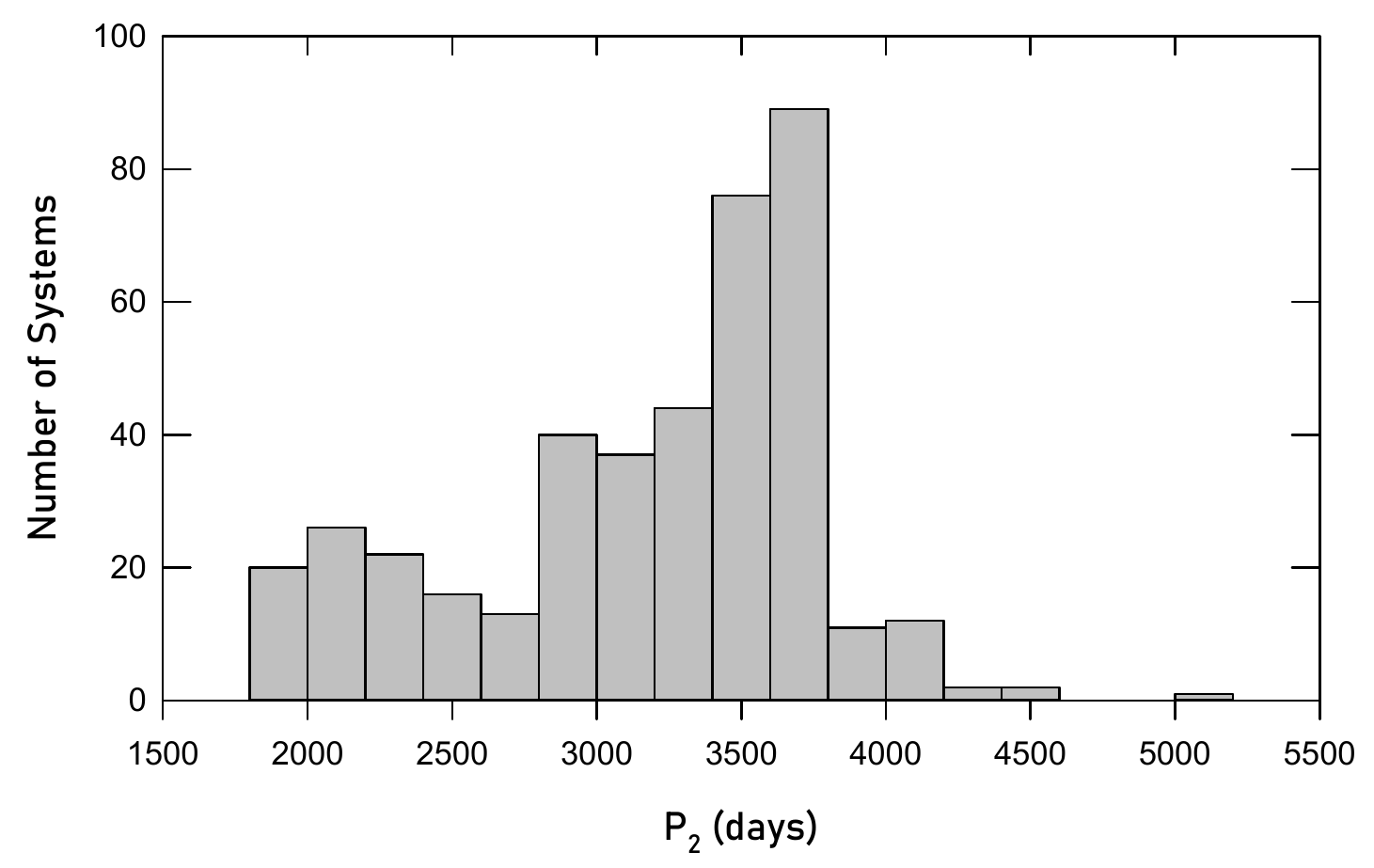} 
\end{tabular}
\caption{Histogram of the outer orbital periods ($P_2$) of the 411 triple candidates.}
\end{center}
\vspace*{0pt}
\end{figure*}


\begin{figure*}[!ht]
\vspace*{0pt}
\begin{center}
\begin{tabular}{c}
\hspace*{0pt}\includegraphics[width=0.6\columnwidth]{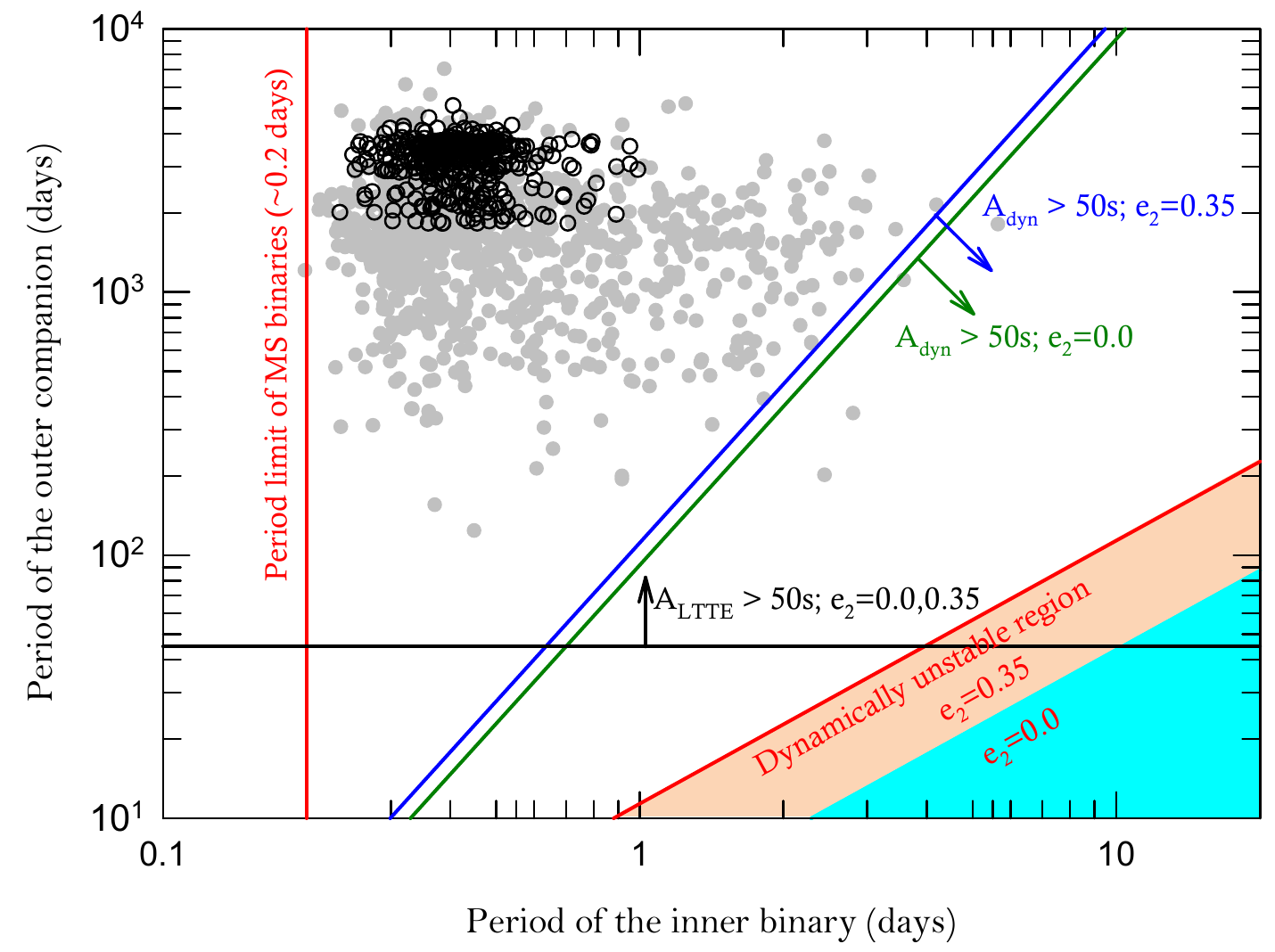} 
\end{tabular}
\caption{Outer triple period ($P_2$) versus the inner binary period ($P_1$) for the 411 triple star candidates (black circles) found in this study. The gray circles represent the 992 triple star candidates in the Galactic Bulge taken from Hajdu et al. (2019). 
The horizontal black, sloped blue (a typical outer eccentricity of $e_2=0.35$) and green ($e_2=0.0$) lines are boundaries of the regions where the amplitudes of the LTTE and dynamical effect may exceed $\sim$50 s or greater in amplitude to be detectable. 
These limits were calculated using the equations (8) and (12) in the paper of Borkovits et al. (2016), where the parameters are adopted as $m_{\rm A}=m_{\rm B}=m_{\rm C}=1$ $M_{\odot}$, $i_2=60^{\circ}$, and $\omega_2=90^{\circ}$.
The dynamically unstable regions for outer eccentricities ($e_2$) of 0.0 (cyan region) and 0.35 (orange region) were calculated using the equation (27) given in the paper by Borkovits (2015), which is rewritten from the expression of Mardling \& Aarseth (2001).
 }
\end{center}
\vspace*{0pt}
\end{figure*}

\clearpage
\begin{figure*}[!ht]
\vspace*{0pt}
\begin{center}
\begin{tabular}{ccc}
\hspace*{0pt}\includegraphics[width=0.7\columnwidth]{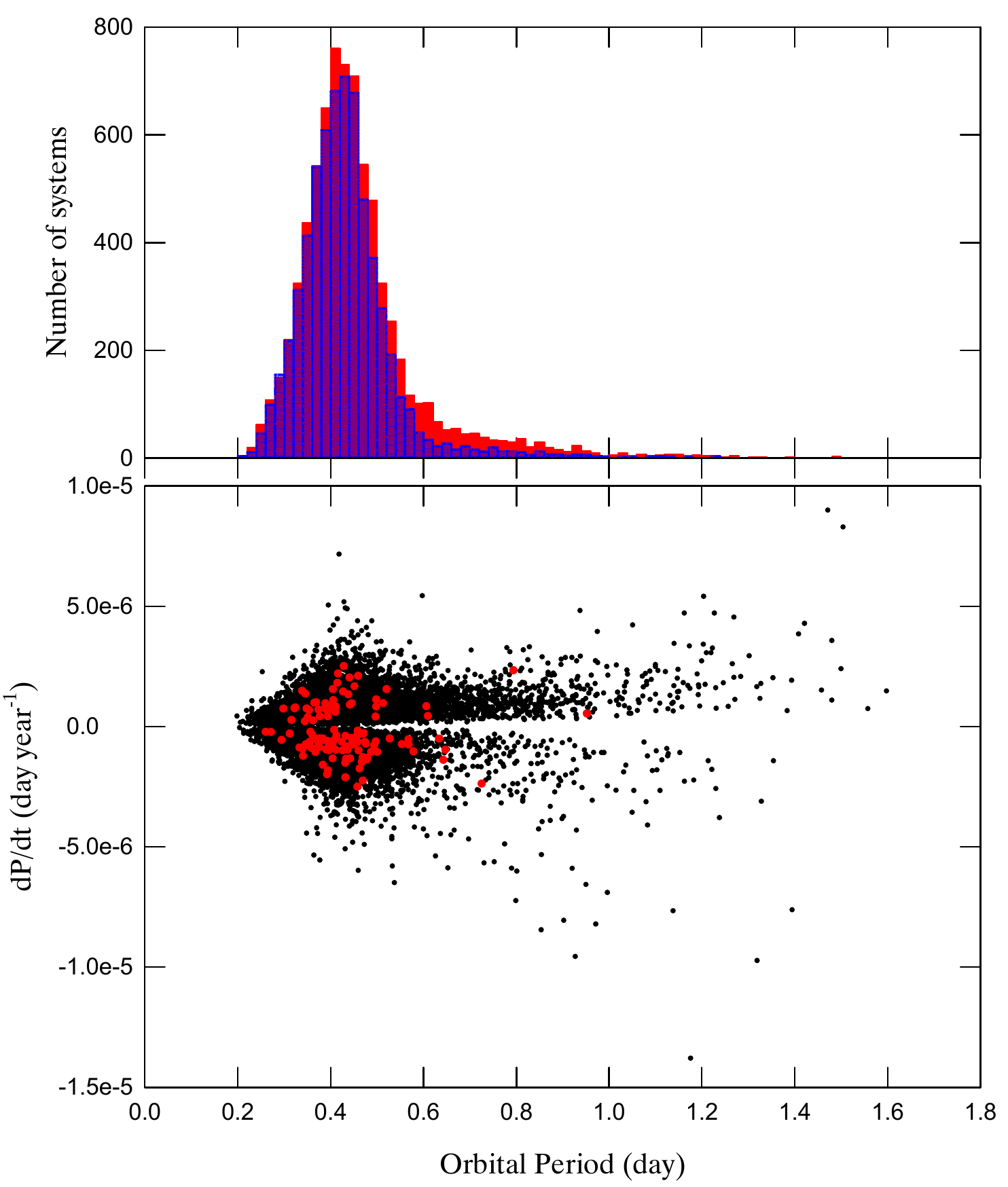} 
\end{tabular}
\caption{The histogram of the orbital period (upper panel) and the period change rate versus the orbital period diagram (lower panel) of the 13,820 CEBs. In the upper panel, the blue and red bars are represent the CEBs with the decreasing and increasing period change rates, respectively. In the lower panel, the black and red circles represent the 13,716 CEBs with parabolic variations and 104 CEBs with parabolic $plus$ sinusoidal variations, respectively. 
}
\end{center}
\vspace*{0pt}
\end{figure*}


\clearpage
\begin{deluxetable}{lccccccrrr}
\tabletypesize{\tiny}
\tablewidth{0pt}
\tablecaption{Period Change Rates of 45 CEBs Overlapped with those of Pietrukowicz et al. (2017)}
\tablehead{
\multicolumn{5}{c}{}  & \multicolumn{3}{c}{This study}  & & \colhead{Pietrukowicz et al. (2017)}              \\                                                                                                                                                                                          
[1.0mm]   \cline{6-8}  \cline{10-10}  \\  [-2.0ex]  
\colhead{Object ID}  &\colhead{R.A.}       &\colhead{Decl.}   & \colhead{$I$}      &\colhead{$V-I$}  &\colhead{$P$}   &\colhead{$T_0$}      &\colhead{$\dot{P}$}                       &  & \colhead{$\dot{P}$} \\                       
\colhead{(OGLE-BLG-)}              &\colhead{(J2000)}    &\colhead{(J2000)} & \colhead{(mag)}    &\colhead{(mag)}  &\colhead{(day)}       &\colhead{(HJD+2450000)}       &\colhead{(day year$^{-1}$)}     &  & \colhead{(day year$^{-1}$)}      
}                                                                                                                                                      
\startdata
ECL-106230  &  17:47:21.97 & $-$34:40:49.3  &  16.056  & 1.330  &  0.4890080(1)   & 7000.4740(2)  &  $ -1.15 \pm 0.02 $ E$-$6  & & $ -8.88 \pm 0.45 $  E$-$7    \\
ECL-114280  &  17:48:28.67 & $-$35:05:09.9  &  16.314  & 1.292  &  0.4530149(3)   & 7000.2223(6)  &  $ -1.78 \pm 0.05 $ E$-$6  & & $ -9.99 \pm 1.64 $  E$-$7    \\
ECL-122558  &  17:49:35.14 & $-$30:48:09.5  &  16.769  & 1.927  &  0.4109427(1)   & 7000.3873(1)  &  $ -3.10 \pm 0.02 $ E$-$7  & & $ -6.79 \pm 1.20 $  E$-$7    \\
ECL-125481  &  17:49:57.46 & $-$29:16:03.0  &  15.070  & 1.611  &  0.3585634(1)   & 7000.0308(1)  &  $  6.51 \pm 0.01 $ E$-$7  & & $  5.88 \pm 0.27 $  E$-$7    \\
ECL-149279  &  17:52:26.29 & $-$32:13:12.7  &  17.688  & 2.043  &  0.3147148(2)   & 7000.2520(5)  &  $ -1.39 \pm 0.04 $ E$-$6  & & $ -2.03 \pm 0.32 $  E$-$6    \\
ECL-153698  &  17:52:48.61 & $-$29:06:42.9  &  16.678  & 1.672  &  0.3866363(1)   & 7000.2255(2)  &  $  1.13 \pm 0.02 $ E$-$6  & & $  1.08 \pm 0.14 $  E$-$6    \\
ECL-157410  &  17:53:07.74 & $-$30:27:52.4  &  15.436  & 1.307  &  0.3634261(1)   & 7000.0597(1)  &  $  1.49 \pm 0.01 $ E$-$6  & & $  1.69 \pm 0.08 $  E$-$6    \\
ECL-159116  &  17:53:16.11 & $-$29:48:59.7  &  15.842  & 1.261  &  0.5873745(3)   & 7000.2863(5)  &  $ -1.87 \pm 0.04 $ E$-$6  & & $ -1.72 \pm 0.06 $  E$-$6    \\
ECL-172659  &  17:54:22.13 & $-$30:01:42.0  &  18.196  & 1.630  &  0.6261682(11)  & 7000.5112(16) &  $ -5.39 \pm 0.17 $ E$-$6  & & $ -5.85 \pm 0.21 $  E$-$6    \\
ECL-174839  &  17:54:34.06 & $-$29:24:38.3  &  16.385  & 1.431  &  0.4612320(1)   & 7000.2364(2)  &  $  7.66 \pm 0.02 $ E$-$7  & & $  9.62 \pm 0.24 $  E$-$7    \\
ECL-181083  &  17:55:07.21 & $-$29:58:17.8  &  16.138  & 1.619  &  0.3493719(1)   & 7000.1897(2)  &  $ -1.78 \pm 0.03 $ E$-$6  & & $ -1.02 \pm 0.04 $  E$-$6    \\
ECL-181311  &  17:55:08.47 & $-$29:33:46.8  &  16.602  & 1.341  &  0.4226041(2)   & 7000.1641(5)  &  $ -1.41 \pm 0.04 $ E$-$6  & & $ -1.43 \pm 0.05 $  E$-$6    \\
ECL-181955  &  17:55:11.75 & $-$30:02:12.1  &  16.212  & 1.419  &  0.4143063(1)   & 7000.0863(1)  &  $  4.45 \pm 0.01 $ E$-$7  & & $  2.62 \pm 0.14 $  E$-$7    \\
ECL-183497  &  17:55:19.37 & $-$32:55:40.9  &  17.803  & 1.105  &  0.5232081(6)   & 7000.4452(10) &  $  1.38 \pm 0.10 $ E$-$6  & & $  3.36 \pm 0.79 $  E$-$6    \\
ECL-184767  &  17:55:25.23 & $-$29:56:15.9  &  17.363  & 1.821  &  0.4732967(4)   & 7000.2416(6)  &  $ -1.01 \pm 0.06 $ E$-$6  & & $ -8.57 \pm 1.18 $  E$-$7    \\
ECL-187430  &  17:55:38.54 & $-$29:17:24.1  &  17.464  & 2.089  &  0.2753081(1)   & 7000.0050(2)  &  $ -3.99 \pm 0.01 $ E$-$7  & & $ -3.18 \pm 0.54 $  E$-$7    \\
ECL-192324  &  17:56:03.39 & $-$29:25:10.6  &  17.785  & 2.144  &  0.4369188(3)   & 7000.3381(6)  &  $  1.71 \pm 0.04 $ E$-$6  & & $  2.05 \pm 0.24 $  E$-$6    \\
ECL-192437  &  17:56:03.96 & $-$29:59:12.6  &  16.808  & 1.627  &  0.4022991(2)   & 7000.0568(3)  &  $ -1.67 \pm 0.04 $ E$-$6  & & $ -1.60 \pm 0.07 $  E$-$6    \\
ECL-192939  &  17:56:06.40 & $-$29:29:21.4  &  18.206  & 2.032  &  0.5318869(8)   & 7000.4773(15) &  $ -4.68 \pm 0.12 $ E$-$6  & & $ -4.40 \pm 0.20 $  E$-$6    \\
ECL-194517  &  17:56:14.16 & $-$29:43:37.7  &  17.159  & 1.580  &  0.4216062(2)   & 7000.2498(5)  &  $  1.35 \pm 0.03 $ E$-$6  & & $  1.58 \pm 0.20 $  E$-$6    \\
ECL-197927  &  17:56:32.09 & $-$29:15:43.0  &  17.211  & 1.761  &  0.5455864(3)   & 7000.4951(5)  &  $  2.13 \pm 0.04 $ E$-$6  & & $  2.26 \pm 0.27 $  E$-$6    \\
ECL-198303  &  17:56:33.75 & $-$30:14:33.7  &  14.734  & 1.197  &  0.4387309(1)   & 7000.1706(1)  &  $ -3.44 \pm 0.01 $ E$-$7  & & $ -3.81 \pm 0.72 $  E$-$7    \\
ECL-199259  &  17:56:38.43 & $-$29:48:45.8  &  13.812  & 0.923  &  0.3892928(1)   & 7000.3713(3)  &  $  8.86 \pm 0.02 $ E$-$7  & & $  1.37 \pm 0.09 $  E$-$6    \\
ECL-200313  &  17:56:43.78 & $-$30:49:34.2  &  17.265  & 1.492  &  0.4791391(1)   & 7000.1185(2)  &  $  9.30 \pm 0.03 $ E$-$7  & & $  1.03 \pm 0.05 $  E$-$6    \\
ECL-200942  &  17:56:46.93 & $-$30:00:57.7  &  18.390  & 1.915  &  0.3583932(3)   & 7000.0824(6)  &  $  1.81 \pm 0.06 $ E$-$6  & & $  2.86 \pm 0.52 $  E$-$6    \\
ECL-201682  &  17:56:50.80 & $-$29:44:49.4  &  17.355  & 1.837  &  0.4973019(3)   & 7000.2404(6)  &  $ -1.21 \pm 0.05 $ E$-$6  & & $ -1.89 \pm 0.29 $  E$-$6    \\
ECL-202706  &  17:56:55.80 & $-$28:39:16.6  &  15.889  & 1.656  &  0.4497383(1)   & 7000.3855(1)  &  $  4.11 \pm 0.01 $ E$-$7  & & $  4.08 \pm 0.71 $  E$-$7    \\
ECL-203756  &  17:57:00.79 & $-$29:39:38.4  &  17.980  & 1.917  &  0.4183556(4)   & 7000.3487(10) &  $  2.10 \pm 0.06 $ E$-$6  & & $  2.14 \pm 0.41 $  E$-$6    \\
ECL-205743  &  17:57:12.10 & $-$29:56:21.2  &  15.745  & 1.394  &  0.3004912(1)   & 7000.0351(1)  &  $  3.05 \pm 0.01 $ E$-$7  & & $  8.24 \pm 0.97 $  E$-$7    \\
ECL-207879  &  17:57:22.78 & $-$28:56:59.0  &  15.064  & 1.474  &  0.3930792(1)   & 7000.0772(1)  &  $ -2.88 \pm 0.01 $ E$-$7  & & $ -5.51 \pm 0.33 $  E$-$7    \\
ECL-212252  &  17:57:44.48 & $-$31:03:40.9  &  18.765  & \dots  &  0.4132157(4)   & 7000.2266(7)  &  $ -3.27 \pm 0.09 $ E$-$6  & & $ -3.88 \pm 0.17 $  E$-$6    \\
ECL-215121  &  17:58:00.27 & $-$29:57:49.4  &  17.125  & 1.867  &  0.3078372(1)   & 7000.0778(1)  &  $ -6.10 \pm 0.01 $ E$-$7  & & $ -5.51 \pm 1.13 $  E$-$7    \\
ECL-220028  &  17:58:24.67 & $-$29:57:42.0  &  16.705  & 1.492  &  0.9207411(6)   & 7000.7851(4)  &  $ -5.91 \pm 0.12 $ E$-$6  & & $ -6.02 \pm 1.10 $  E$-$6    \\
ECL-233754  &  17:59:37.56 & $-$29:12:27.8  &  16.919  & 1.298  &  0.5702720(4)   & 7000.5608(5)  &  $  3.51 \pm 0.07 $ E$-$6  & & $  1.84 \pm 0.09 $  E$-$6    \\
ECL-233821  &  17:59:37.95 & $-$28:50:22.4  &  15.276  & 1.220  &  0.3599062(2)   & 7000.3556(4)  &  $ -1.19 \pm 0.03 $ E$-$6  & & $ -1.59 \pm 0.04 $  E$-$6    \\
ECL-279326  &  18:03:46.02 & $-$29:47:40.5  &  15.765  & 1.519  &  0.2277294(1)   & 7000.1718(1)  &  $ -3.49 \pm 0.01 $ E$-$7  & & $ -3.29 \pm 0.24 $  E$-$7    \\
ECL-280295  &  18:03:51.13 & $-$27:54:12.3  &  16.145  & 1.356  &  0.4842285(1)   & 7000.4589(1)  &  $ -1.02 \pm 0.01 $ E$-$6  & & $ -1.10 \pm 0.10 $  E$-$6    \\
ECL-282055  &  18:04:00.72 & $-$28:39:33.7  &  17.620  & 1.560  &  0.4828396(4)   & 7000.2603(5)  &  $  2.93 \pm 0.07 $ E$-$6  & & $  2.04 \pm 0.14 $  E$-$6    \\
ECL-284531  &  18:04:14.75 & $-$29:52:19.9  &  13.598  & 0.883  &  0.4355595(1)   & 7000.2722(1)  &  $  5.58 \pm 0.01 $ E$-$7  & & $  5.17 \pm 0.21 $  E$-$7    \\
ECL-294795  &  18:05:10.60 & $-$29:21:03.9  &  13.019  & 1.000  &  0.2936554(1)   & 7000.1750(2)  &  $ -3.35 \pm 0.01 $ E$-$7  & & $ -4.84 \pm 0.41 $  E$-$7    \\
ECL-306303  &  18:06:16.05 & $-$29:00:46.8  &  17.850  & 1.099  &  0.4155388(5)   & 7000.0956(9)  &  $  2.36 \pm 0.08 $ E$-$6  & & $  3.31 \pm 0.59 $  E$-$6    \\
ECL-311367  &  18:06:47.01 & $-$30:47:49.1  &  17.965  & 1.497  &  0.4202339(6)   & 7000.3097(14) &  $  3.57 \pm 0.11 $ E$-$6  & & $  4.44 \pm 0.80 $  E$-$6    \\
ECL-317072  &  18:07:22.35 & $-$29:42:13.1  &  15.582  & 1.173  &  0.3537176(1)   & 7000.2037(2)  &  $  6.53 \pm 0.02 $ E$-$7  & & $  7.28 \pm 0.82 $  E$-$7    \\
ECL-317088  &  18:07:22.46 & $-$29:18:39.2  &  17.678  & 1.297  &  0.4874658(5)   & 7000.1490(8)  &  $  2.56 \pm 0.09 $ E$-$6  & & $  3.47 \pm 0.69 $  E$-$6    \\
ECL-351804  &  18:11:16.73 & $-$25:42:55.3  &  16.684  & 1.426  &  0.3853975(2)   & 7000.3813(4)  &  $ -2.33 \pm 0.03 $ E$-$6  & & $ -2.12 \pm 0.26 $  E$-$6    \\
\enddata
 \begin{list}{}{}
 \item Note: The coordinates, $I$, and $V-I$ are taken from the OGLE-IV catalogue by Soszy\'nski et al. (2016).   \\[-2.0ex]  
 \end{list}
\end{deluxetable}


\clearpage
\begin{deluxetable}{lccccccc}
\tabletypesize{\scriptsize}
\tablewidth{0pt}
\tablecaption{The Secular Period Change Rates of 13,716 CEBs}
\tablehead{
\colhead{Object ID}  &\colhead{R.A.}       &\colhead{Decl.}   & \colhead{$I$}      &\colhead{$V-I$}  &\colhead{$P$}   &\colhead{$T_0$}      &\colhead{$\dot{P}$}     \\                       
\colhead{(OGLE-BLG-)}     &\colhead{(J2000)}    &\colhead{(J2000)} & \colhead{(mag)}    &\colhead{(mag)}  &\colhead{(day)}       &\colhead{(HJD+2450000)}       &\colhead{(day year$^{-1}$)} 
}                                         
\startdata                                                                                                                                             
ECL-169991  &   17:54:09.00  & $-$32:05:55.0   &      17.104   &   1.965  &   1.1757013(27)   &   2457000.8180(15)    &   $ -1.38 \pm 0.06 $  E$-$5  \\
ECL-052921  &   17:38:43.52  & $-$22:36:37.2   &      16.425   &   1.672  &   1.3190508(36)   &   2457000.3542(22)    &   $ -9.74 \pm 0.54 $  E$-$6  \\
ECL-218952  &   17:58:19.54  & $-$28:08:21.9   &      16.662   &   2.211  &   0.9275454(7)    &   2457000.9101(5)     &   $ -9.57 \pm 0.18 $  E$-$6  \\
ECL-272254  &   18:03:07.25  & $-$28:15:19.0   &      15.549   &   1.168  &   0.8535252(4)    &   2457000.8072(2)     &   $ -8.46 \pm 0.08 $  E$-$6  \\
ECL-151250  &   17:52:36.28  & $-$30:42:53.2   &      17.127   &   2.175  &   0.9717356(9)    &   2457000.7534(6)     &   $ -8.22 \pm 0.19 $  E$-$6  \\
ECL-122763  &   17:49:36.96  & $-$34:40:44.4   &      17.252   &   1.560  &   0.9023381(25)   &   2457000.1415(25)    &   $ -8.07 \pm 0.43 $  E$-$6  \\
ECL-339477  &   18:09:50.66  & $-$25:58:29.2   &      16.812   &   1.535  &   1.1381204(25)   &   2457001.1215(15)    &   $ -7.67 \pm 0.44 $  E$-$6  \\
\dots       &     &    &    &     &     &      &      \\
\enddata
 \begin{list}{}{}
 \item Note: The coordinates, $I$, and $V-I$ are taken from the OGLE-IV catalogue by Soszy\'nski et al. (2016).   \\[-2.0ex]  
 \item (This table is published in its entirety in the machine-readable format.
      A portion is shown here for guidance regarding its form and content.)
 \end{list}
\end{deluxetable}


\clearpage
\begin{deluxetable}{lccccccccc}
\tabletypesize{\scriptsize}
\tablewidth{0pt}
\tablecaption{Results of Sine-curve fit for the 307 CEBs with a Sinusodal Period Change}
\tablehead{
\multicolumn{5}{c}{}  & \multicolumn{3}{c}{This study}  & &  \colhead{The others}             \\                                                                                                                                                                                          
[1.0mm]   \cline{6-8}   \cline{10-10}     \\  [-2.0ex]  
\colhead{OGLE-BLG-}  &\colhead{R.A.}  &\colhead{Decl.} & \colhead{$I$} &\colhead{$V-I$}  &\colhead{$P$}  &\colhead{$T_0$} &\colhead{$P_2$} & &\colhead{$P_2$}                    \\                       
\colhead{}   &\colhead{(J2000)}    &\colhead{(J2000)} & \colhead{(mag)}    &\colhead{(mag)}  &\colhead{(day)}  &\colhead{(HJD+2450000)}     &\colhead{(year)}    & & \colhead{(year)}      
}                                                                                                                                                      
\startdata                       
\dots       &               &                  &           &           &                   &                     &                 &  &              \\
ECL-124673  &  17:49:51.64  &  $-$30:06:25.3   &  16.771   &   1.650   &    0.3540138(1)   &   2457000.0105(1)   &      8.3(4)     &  &  4.5(1)$^{\rm b}$           \\
ECL-125353  &  17:49:56.56  &  $-$34:58:02.1   &  17.953   &   1.607   &    0.5121666(3)   &   2457000.1523(12)  &     10.2(5)     &  &              \\
ECL-125525  &  17:49:57.66  &  $-$34:39:54.4   &  14.800   &   1.212   &    0.4120029(1)   &   2457000.2652(3)   &     10.9(4)     &  &  15$^{\rm a}$             \\
ECL-127170  &  17:50:09.95  &  $-$29:44:30.3   &  17.778   &   1.902   &    0.2915032(1)   &   2457000.2528(2)   &     11.0(3)     &  &              \\
ECL-127614  &  17:50:13.22  &  $-$34:57:38.6   &  17.532   &   1.470   &    0.4789311(2)   &   2457000.0820(8)   &      5.5(1)     &  &              \\
ECL-127683  &  17:50:13.66  &  $-$34:42:15.3   &  18.882   &   1.462   &    0.3518262(1)   &   2457000.2028(10)  &      5.8(1)     &  &              \\
ECL-127744  &  17:50:14.07  &  $-$30:06:22.4   &  15.017   &   1.741   &    0.8920668(1)   &   2457000.6614(2)   &      5.4(3)     &  &  6.5$^{\rm a}$, 4.0(2)$^{\rm b}$              \\
\dots       &               &                  &           &           &                   &                     &                 &  &              \\
\enddata                                                                                                            
 \begin{list}{}{}
 \item Note: The coordinates, $I$, and $V-I$ are taken from the OGLE-IV catalogue by Soszy\'nski et al. (2016). 
 \item $^{\rm a}$ Value taken from Pietrukowicz et al. (2017).
 \item $^{\rm b}$ Value taken from Hajdu et al. (2019).
 \item (This table is published in its entirety in the machine-readable format.
      A portion is shown here for guidance regarding its form and content.)
\end{list}
\end{deluxetable}

\clearpage
\begin{deluxetable}{lcccclcrccc}
\tabletypesize{\tiny} 
\tablewidth{0pt}
\tablecaption{The fitted parameters for the parabolic $plus$ sinusoidal ephemeris of 104 Contact EBs}
\tablehead{
\multicolumn{5}{c}{}  & \multicolumn{4}{c}{This study}   & \colhead{}              \\                                                                                                                                                                                          
[1.0mm]   \cline{6-9} \\ 
\colhead{OGLE-BLG-}  &\colhead{R.A.}       &\colhead{Decl.}   & \colhead{$I$}      &\colhead{$V-I$}  &\colhead{$P$}   &\colhead{$T_0$} &\colhead{$\dot{P}$}   &\colhead{$P_2$}  &  & \colhead{$P_2^{*}$} \\                       
\colhead{}              &\colhead{(J2000)}    &\colhead{(J2000)} & \colhead{(mag)}    &\colhead{(mag)}  &\colhead{(day)}       &\colhead{(HJD+2450000)}       &\colhead{(day year$^{-1}$)}  &\colhead{(year)}   &   & \colhead{(year)}      
}                                                                                                                                                      
\startdata                                                                                                         
\dots       &              &               &          &        &                &                &                         &             &  &       \\                                    
ECL-147315  & 17:52:15.66  & $-$33:20:21.7 &  15.744  &  1.341 & 0.3739789(1)   & 7000.1115(2)   & $ -6.69 \pm 1.94$ E$-$7 &    9.9(2)   &  &       \\
ECL-148291  & 17:52:21.01  & $-$29:46:59.3 &  16.952  &  1.612 & 0.3388654(2)   & 7000.2649(4)   & $  1.46 \pm 0.21$ E$-$6 &    9.8(1)   &  &  21$^{\rm a}$, 10.5(1.1)$^{\rm b}$ \\
ECL-151538  & 17:52:37.82  & $-$31:20:26.7 &  18.368  &  2.395 & 0.5786571(4)   & 7000.3272(4)   & $ -1.05 \pm 0.31$ E$-$6 &    9.8(2)   &  &       \\
ECL-157106  & 17:53:06.12  & $-$29:58:36.5 &  17.794  &  1.826 & 0.3224295(2)   & 7000.1954(6)   & $  7.57 \pm 2.04$ E$-$7 &    9.8(1)   &  &       \\
ECL-159957  & 17:53:20.30  & $-$29:24:57.3 &  16.696  &  1.507 & 0.3455664(2)   & 7000.1874(5)   & $  1.36 \pm 0.22$ E$-$6 &   10.0(2)   &  &       \\
ECL-163882  & 17:53:39.93  & $-$33:53:45.1 &  17.241  &  1.481 & 0.4790803(3)   & 7000.2601(5)   & $ -1.32 \pm 0.25$ E$-$6 &    9.8(2)   &  &       \\
ECL-166853  & 17:53:54.55  & $-$33:11:01.9 &  15.796  &  1.535 & 0.3405647(2)   & 7000.2570(4)   & $ -1.22 \pm 0.17$ E$-$6 &    9.6(1)   &  &       \\    
\dots       &              &               &          &        &                &                &                         &             &  &       \\                                    
\enddata
 \begin{list}{}{}
 \item Note: The coordinates, $I$, and $V-I$ are taken from the OGLE-IV catalogue by Soszy\'nski et al. (2016). 
 \item $^{\rm a}$ Value taken from Pietrukowicz et al. (2017).
 \item $^{\rm b}$ Value taken from Hajdu et al. (2019).
 \item (This table is published in its entirety in the machine-readable format.
      A portion is shown here for guidance regarding its form and content.) 
 \end{list}
\end{deluxetable}


\begin{thebibliography}{}
\bibitem[Applegate(1992)]{1992ApJ...385..621A} Applegate, J.~H.\ 1992, \apj, 385, 621
\bibitem[Benjamini \& Hochberg(1995)]{1995JRSS.57,289} Benjamini, Y., \& Hochberg, Y., 1995, Journal of the Royal Statistical Society, Series B 57, 289
\bibitem[Borkovits et al.(2015)]{2015MNRAS.448..946B} Borkovits, T., Rappaport, S., Hajdu, T., et al.\ 2015, \mnras, 448, 946
\bibitem[Borkovits et al.(2016)]{2016MNRAS.455.4136B} Borkovits, T., Hajdu, T., Sztakovics, J., et al.\ 2016, \mnras, 455, 4136
 \bibitem[Brown et al.(2002)]{2002IAUC.7785....1B} Brown, N.~J., Waagen, E.~O., Scovil, C., et al.\ 2002, \iaucirc 7785, 1
4136
\bibitem[Eggleton(2012)]{2012JASS...29..145E} Eggleton, P.~P.\ 2012, Journal of Astronomy and Space Sciences, 29, 145
\bibitem[Ferreira et al.(2019)]{2019MNRAS.486.1220F} Ferreira, T., Saito, R.~K., Minniti, D., et al.\ 2019, \mnras, 486, 1220
\bibitem[Gazeas et al.(2021)]{2021MNRAS.502.2879G} Gazeas, K.~D., Loukaidou, G.~A., Niarchos, P.~G., et al.\ 2021, \mnras, 502, 2879  
\bibitem[Hajdu et al.(2019)]{2019MNRAS.485.2562H} Hajdu, T., Borkovits, T., Forg{\'a}cs-Dajka, E., et al.\ 2019, \mnras, 485, 2562
\bibitem[Hayashi et al.(1994)]{1994IAUC.5942....1H} Hayashi, S.~S., Yamamoto, M., \& Hirosawa, K.\ 1994, \iaucirc 5942, 1
\bibitem[Hong et al.(2015)]{2015AJ....150....1H} Hong, K., Kim, S.-L., Lee, J.~W., et al.\ 2015, \aj, 150, 1
\bibitem[Hong et al.(2016)]{2016MNRAS.460..650H} Hong, K., Lee, J.~W., Kim, S.-L., Koo, J.-R., \& Lee, C.-U.\ 2016, \mnras, 460, 650
\bibitem[Hong et al.(2019)]{2019AJ....158..185H} Hong, K., Lee, J.~W., Kim, S.-L., et al.\ 2019, \aj, 158, 185
\bibitem[Howitt et al.(2020)]{2020MNRAS.492.3229H} Howitt, G., Stevenson, S., Vigna-G{\'o}mez, A., et al.\ 2020, \mnras, 492, 3229
\bibitem[Ihaka \& Gentleman(1996)]{1996JCGS.5.299} Ihaka R., \& Gentleman R. 1996, Journal of Computational and Graphical Statistics, 5, 299
\bibitem[Irwin(1952)]{1952ApJ...116..211I} Irwin, J.~B.\ 1952, \apj, 116, 211 
\bibitem[Irwin(1959)]{1959AJ.....64..149I} Irwin, J.~B.\ 1959, \aj, 64, 149
\bibitem[Ivanova et al.(2013)]{2013Sci...339..433I} Ivanova, N., Justham, S., Avendano Nandez, J.~L., et al.\ 2013, Science, 339, 433
\bibitem[Kashi(2018)]{2018Galax...6...82K} Kashi, A.\ 2018, Galaxies, 6, 82
\bibitem[Kass \& Raftery(1995)]{1995JASA.90.773} Kass, R. E., \& Raftery, A. E. 1995, Journal of the American Statistical Association, 90, 773
\bibitem[Kochanek et al.(2014)]{2014MNRAS.443.1319K} Kochanek, C.~S., Adams, S.~M., \& Belczynski, K.\ 2014, \mnras, 443, 1319
\bibitem[Kovacs et al.(2019)]{2019A&A...631A.126K} Kovacs, G., Hartman, J.~D., \& Bakos, G. {\'A}.\ 2019, \aap, 631, A126
\bibitem[Kubiak et al.(2006)]{2006AcA....56..253K} Kubiak, M., Udalski, A., \& Szymanski, M.~K.\ 2006, \actaa, 56, 253
\bibitem[Kulkarni et al.(2007)]{2007Natur.447..458K} Kulkarni, S.~R., Ofek, E.~O., Rau, A., et al.\ 2007, \nat, 447, 458
\bibitem[Lanza et al.(1998)]{1998MNRAS.296..893L} Lanza, A.~F., Rodono, M., \& Rosner, R.\ 1998, \mnras, 296, 893
\bibitem[Lee et al.(2014)]{2014AJ....147...91L} Lee, J.~W., Park, J.-H., Hong, K., et al.\ 2014, \aj, 147, 91
\bibitem[Lucy(1968)]{1968ApJ...151.1123L} Lucy, L.~B.\ 1968, \apj, 151, 1123
\bibitem[Lucy(1976)]{1976ApJ...205..208L} Lucy, L.~B.\ 1976, \apj, 205, 208
\bibitem[Maceroni \& van't Veer(1994)]{1994A&A...289..871M} Maceroni, C. \& van't Veer, F.\ 1994, \aap, 289, 871
\bibitem[Maceroni \& van't Veer(1996)]{1996A&A...311..523M} Maceroni, C. \& van't Veer, F.\ 1996, \aap, 311, 523
\bibitem[Mardling \& Aarseth(2001)]{2001MNRAS.321..398M} Mardling, R.~A. \& Aarseth, S.~J.\ 2001, \mnras, 321, 398
\bibitem[Mochnacki(1981)]{1981ApJ...245..650M} Mochnacki, S.~W.\ 1981, \apj, 245, 650
\bibitem[Molnar et al.(2017)]{2017ApJ...840....1M} Molnar, L.~A., Van Noord, D.~M., Kinemuchi, K., et al.\ 2017, \apj, 840, 1 
\bibitem[Nakano et al.(2008)]{2008CBET.1496....1N} Nakano, S., Nishiyama, K., Kabashima, F., et al.\ 2008, Central Bureau Electronic Telegrams, 1496, 1
\bibitem[Nandez et al.(2014)]{2014ApJ...786...39N} Nandez, J.~L.~A., Ivanova, N., \& Lombardi, J.~C.\ 2014, \apj, 786, 39
\bibitem[Pastorello et al.(2019)]{2019A&A...630A..75P} Pastorello, A., Mason, E., Taubenberger, S., et al.\ 2019, \aap, 630, A75  
\bibitem[Paczy{\'n}ski(1967)]{1967AcA....17..287P} Paczy{\'n}ski, B.\ 1967, \actaa, 17, 287
\bibitem[Paczy{\'n}ski et al.(2006)]{2006MNRAS.368.1311P} Paczy{\'n}ski, B., Szczygie{\l}, D.~M., Pilecki, B., et al.\ 2006, \mnras, 368, 1311
\bibitem[Pietrukowicz et al.(2017)]{2017AcA....67..115P} Pietrukowicz, P., Soszy{\'n}ski, I., Udalski, A., et al.\ 2017, \actaa, 67, 115
\bibitem[Prendergast \& Taam(1974)]{1974ApJ...189..125P} Prendergast, K.~H. \& Taam, R.~E.\ 1974, \apj, 189, 125
\bibitem[Press et al.(1992)]{1992nrfa.book.....P} Press, W.~H., Teukolsky, S.~A., Vetterling, W.~T., et al.\ 1992, Numerical Recipes (Cambridge: Cambridge Univ. Press)
\bibitem[Rappaport et al.(1983)]{1983ApJ...275..713R} Rappaport, S., Verbunt, F., \& Joss, P.~C.\ 1983, \apj, 275, 713
\bibitem[Socia et al.(2018)]{2018ApJ...864L..32S} Socia, Q.~J., Welsh, W.~F., Short, D.~R., et al.\ 2018, \apjl, 864, L32
\bibitem[Soker \& Tylenda(2003)]{2003ApJ...582L.105S} Soker, N., \& Tylenda, R.\ 2003, \apjl, 582, L105
\bibitem[Soszy{\'n}ski et al.(2016)]{2016AcA....66..405S} Soszy{\'n}ski, I., Pawlak, M., Pietrukowicz, P., et al.\ 2016, \actaa, 66, 405
\bibitem[St{\c{e}}pie{\'n}(2006)]{2006AcA....56..199S}  St{\c{e}}pie{\'n}, K.\ 2006, \actaa, 56, 199
\bibitem[Tylenda et al.(2011)]{2011A&A...528A.114T} Tylenda, R., Hajduk, M., Kami{\'n}ski, T., et al.\ 2011, \aap, 528, A114
\bibitem[Tylenda et al.(2013)]{2013A&A...555A..16T} Tylenda, R., Kami{\'n}ski, T., Udalski, A., et al.\ 2013, \aap, 555, A16
\bibitem[Udalski(2003)]{2003AcA....53..291U} Udalski, A.\ 2003, \actaa, 53, 291
\bibitem[Udalski et al.(1992)]{1992AcA....42..253U} Udalski, A., Szymanski, M., Kaluzny, J., Kubiak, M., \& Mateo, M.\ 1992, \actaa, 42, 253
\bibitem[Udalski et al.(2008)]{2008AcA....58...69U} Udalski, A., Szymanski, M.~K., Soszynski, I., et al.\ 2008, \actaa, 58, 69
\bibitem[Udalski et al.(2015)]{2015AcA....65....1U} Udalski, A., Szyma{\'n}ski, M.~K., \& Szyma{\'n}ski, G.\ 2015, \actaa, 65, 1
\bibitem[Van Hamme \& Wilson(2007)]{2007ApJ...661.1129V} Van Hamme, W., \& Wilson, R.~E.\ 2007, \apj, 661, 1129
\bibitem[van't Veer(1979)]{1979A&A....80..287V} van't Veer, F.\ 1979, \aap, 80, 287
\bibitem[van't Veer \& Maceroni(1989)]{1989A&A...220..128V} van't Veer, F. \& Maceroni, C.\ 1989, \aap, 220, 128
\bibitem[Wagenmakers(2007)]{2007PBR.14.779} Wagenmakers, E.-J. 2007, Psychonomic Bulletin \& Review, 14, 779
\bibitem[Webbink(1976)]{1976ApJ...209..829W} Webbink, R.~F.\ 1976, \apj, 209, 829
\bibitem[Webbink(2003)]{2003ASPC..293...76W} Webbink, R.~F.\ 2003, 3D Stellar Evolution, 293, 76
\bibitem[Wilson \& Devinney(1971)]{1971ApJ...166..605W} Wilson, R.~E., \& Devinney, E.~J.\ 1971, \apj, 166, 605
\end{thebibliography}
\end{document}